\documentclass[journal]{IEEEtran}
\ifCLASSINFOpdf
   \usepackage[pdftex]{graphicx}
\else
\fi
\usepackage[cmex10]{amsmath}
\usepackage{amssymb}
\usepackage{array}
\begin{document}
%
\title{Enhancing Heralding Efficiency and Biphoton Rate in Type-I Spontaneous Parametric Down-Conversion}
%
%
%

\author{Hannah~E.~Guilbert,~\IEEEmembership{Student Member,~IEEE,}
                and~Daniel~J.~Gauthier,~\IEEEmembership{Member,~IEEE, Fellow,~OSA}
\thanks{H. Guilbert and D.~J.~Gauthier are with the Department
of Physics and the Fitzpatrick Institute of Photonics, Duke University, Durham,
NC, 27708 USA (e-mail: heg3@duke.edu, gauthier@phy.duke.edu).}
}

%
%

\markboth{Journal of Selected Topics in Quantum Electronics}%
{Guilbert \MakeLowercase{\textit{et al.}}: Bare Demo of IEEEtran.cls for Journals}
%



\maketitle

\begin{abstract}
The nonlinear optical process of spontaneous parametric down-conversion (SPDC) is widely studied for applications in quantum information science due to its ability to produce 
two photons that can be entangled in many degrees of freedom. For applications in quantum communication, two metrics of this process are particularly important: heralding 
efficiency and total joint rate. Here, we derive expressions for both quantities for a variety of different beam geometries and frequencies. We pay specific attention to the 
spectrum of both  biphotons and individual photons. We reveal the underlying mechanisms responsible for the spectral shape and show they differ for different geometries and 
frequencies. We then use these spectra to calculate heralding efficiency and joint count rate and examine how each of these metrics changes with different geometries, frequencies,
 and spectral filtering and beam parameters. Interestingly, we find very high heralding efficiencies are achievable for collinear geometries, while noncollinear geometries require 
spectral filtering for high heralding efficiency, ultimately limiting the joint count rate. We also find that the the spectrum is narrower in nondegenerate cases, leading to lower 
joint count rates and higher heralding efficiency in the noncollinear case. In addition to the theory, we verify selected predictions with experimental results. 
\end{abstract}


%
\IEEEpeerreviewmaketitle

\section{Introduction}
Spontaneous parametric down-conversion (SPDC) is a nonlinear optical process where a photon from a pump beam passing through a transparent crystal spontaneously splits into two  
 daughter photons.  The SPDC process has received continued interest because the daughter photons can be quantum mechanically entangled in the photon's degrees of freedom under  
appropriate conditions \cite{Shih}.  The entanglement quality can be very high, enabling a wide range of experiments, from testing the foundations of quantum mechanics to 
applications in quantum communication, for example.  Given that the SPDC process is an important resource for quantum information science, several groups have explored methods
 for optimizing different performance metrics of SPDC sources.  The range of accessible experimental parameters is quite large, so many of these studies restrict their analysis 
to a subset of parameter space or resort to numerical simulations, which makes it difficult to easily generalize the results of these studies to different situations.

In greater detail, the SPDC process most often involves the interaction of a pump beam (frequency $\omega_p$, propagation wavevector $\vec{k}_p$), and two generated beams, 
often called signal ($\omega_s$, $\vec{k}_p$) and idler ($\omega_i$, $\vec{k}_i$) beams, propagating through a transparent birefringent crystal characterized by a second-order 
nonlinear optical susceptibility.  Conservation of energy requires that $\omega_p=\omega_s+ \omega_i$, and the SPDC process occurs most efficiently when momentum is conserved
 among the incident and generated beams - often referred to as phase matching - so that $\vec{k}_p=\vec{k}_s+\vec{k}_p$.  In some situations, it is desirable to spatially 
separate the signal and idler beams and couple them into single-mode optical fibers for subsequent processing and transmission. 

\begin{figure}
 \begin{center}
 \includegraphics[scale=0.4]{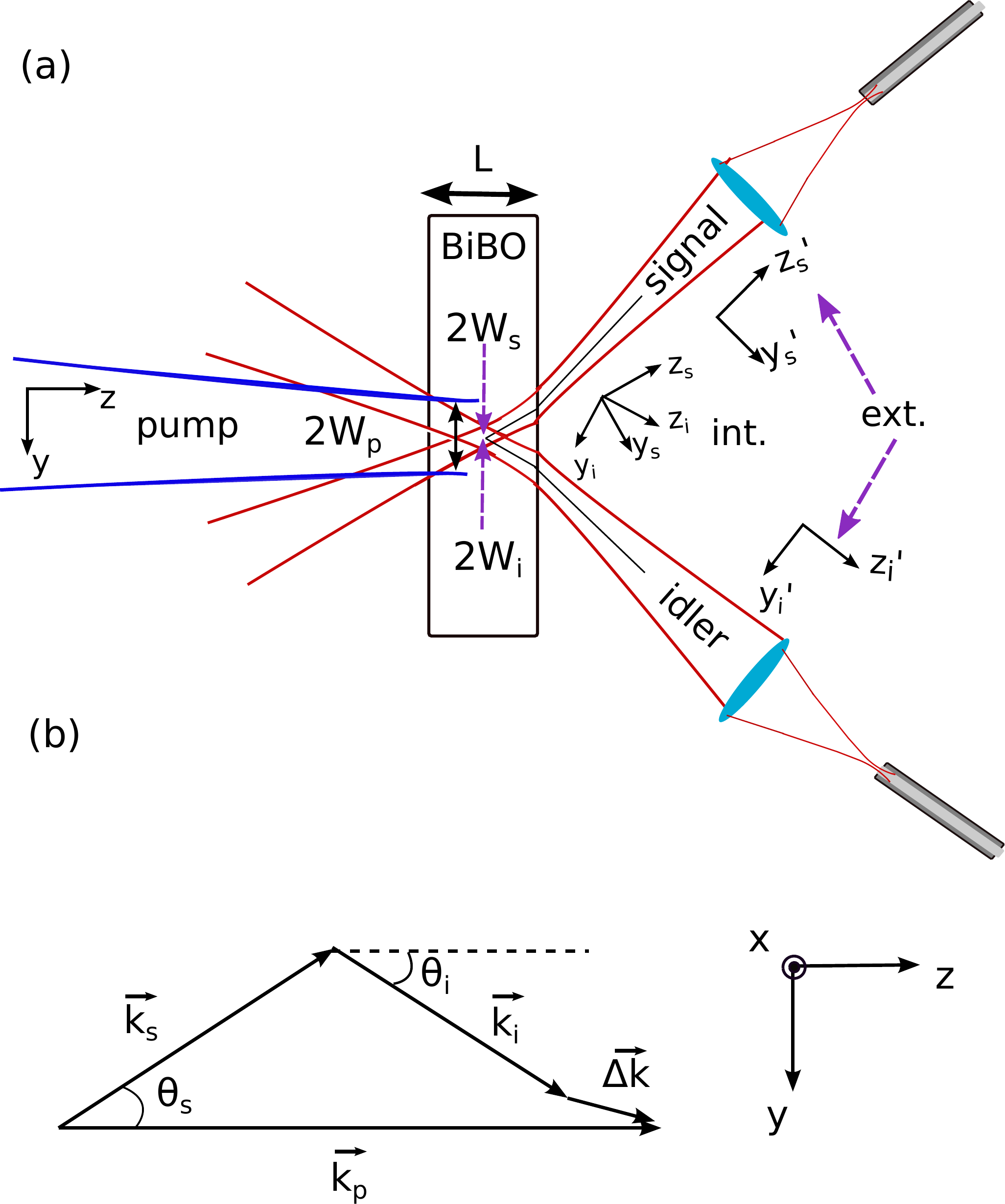}
\end{center}
\caption{\textbf{SPDC interaction geometry} (a) A pump beam (blue) is focused at the center of a nonlinear optical crystal of length $L$. 
 The fixed target modes (red) for the signal (idler) beams are defined by single-mode optical fibers, which are back-imaged to the center of the crystal.
 (b) Relation among the beam wavevectors and the phase mismatch. }
\label{pumpgeo}
\end{figure}

In this paper, we consider a SPDC source consisting of a thin nonlinear optical crystal using the so-called Type-I interaction.  Here, the all the beams are linearly polarized, 
the signal and idler beams are co-polarized, and the pump beam polarization is orthogonal to the generated beams.  Phase matching is obtained by tuning the angle of the beams with 
respect to the crystal optic axes.  In our study, we assume that the pump beam is in the fundamental Gaussian spatial mode, and the signal and idler beams are emitted collinearly 
or at a small angle with respect to the pump beam, shown schematically in Fig.~\ref{pumpgeo}.  This configuration is a basic building block for realizing a bright source of 
polarization-entangled photons \cite{Kwiat95, Kwiat99}, for example.  Subsequently, the generated beams are coupled into single-mode fibers, spectrally filtered, and directed 
to photon-counting detectors.  

We focus on two metrics of the SPDC source: the \textit{joint rate} (also known as the coincidence or biphoton rate) $R$, which is the rate of correlated photon events
 in the signal and idler beam paths, and the symmetric \textit{heralding efficiency} $\eta=R/\sqrt{R_s R_i}$ where $R_s$ ($R_i$) is the rate of single photon detections 
in the signal (idler) path.  Optimizing the joint rate is crucial for quantum key distribution systems, for example, because the secure key rate is proportional to 
$R$ \cite{RevModPhysGisin}.

The heralding efficiency takes on a value less than one when a photon is detected in the signal path, for example, but the correlated photon does not appear in the idler path.  
 Enhancing $\eta$ is crucial for a number of applications that require a minimum value, including detection loophole-free tests of Bell's inequality ($\eta \geq 66\%$) 
\cite{loopholefree, bradprl, zeilenger}, one-sided device-independent QKD (DI-QKD) ($\eta \geq 66\%$) \cite{DIQKD}, and three-party quantum communication \cite{Pingpong1, Pingpong2} 
($\eta \geq 60\%$).  Also,  high heralding efficiency decreases the quantum bit error rate and hence increases the error correction efficiency for point-to-point quantum key 
distribution systems.   

Both the joint rate and the heralding efficiency depend on most of the accessible parameters and there often exists a trade-off between the two: obtaining high $\eta$ often entails a 
reduction in $R$ \cite{Bennink}.  In particular, we find that geometry (collinear or slightly noncollinear), frequency of the generated photons, beam spot sizes, and spectral 
filtering can all can have a important effect on $R$ and $\eta$.  We motivate and give a physical explanation for why high $R$ and $\eta$ can be obtained in some configurations 
and not in others by investigating how the accessible parameters affect both the joint and and signal/idler spectra.  

We find that $\eta > 97\%$ can be obtained for the case of collinear SPDC when the frequencies of the signal and idler beams do not overlap so that they can be spatially separated 
with a dichroic mirror.   Furthermore, we show $R$ and $\eta$ increase with overall decrease in mode waist sizes of pump, signal, and idler modes. For noncollinear SPDC, high $\eta$
 can only be obtained with spectral filtering, which decrease $R$.  We verify selected predictions in an experiment where we pump a thin BiB$_3$O$_6$ (BiBO) crystal pumped by an 
ultraviolet modelocked laser.  

The paper is organized as follows: In Sec.~II we derive both the joint spectrum and the singles spectrum of the SPDC photons and show how the dependence on the phase mismatch makes 
spectra different for different geometries and waists. In Sec.~III we discuss the heralding efficiency and joint count rate in both spectrally filtered and unfiltered cases and
 discuss its dependence on overall scaling of the mode waists as well as relative scaling of mode waists. In Sec.~IV we discuss several parameters that optimize heralding efficiency
 for all cases and briefly discuss how to optimize both heralding efficiency and joint counts for each geometry. In Sec.~V we present  our experimental setup, discuss our results, 
and compare them to our theoretical findings and in Sec.~VI we conclude.

\section{Formalism for Predicting the Joint and Singles Spectral Rates}

The procedure for predicting the joint and singles spectral rates for SPDC into single-mode is well established and can be found across several studies.  In our work, we 
closely follow and use similar notation for the joint spectral rate as described in Ling \textit{et al.} \cite{Ling}, who consider noncollinear SPDC, but do not predict the 
singles spectra or counts, which is needed to predict $\eta$.\footnote{We note that Eq.~(32) of Ling \textit{et al.} is in error, limiting the remainder of their work to 
Type-II SPDC or to non-degenerate Type-I SPDC.  We do not use any results from Ling \textit{et al.} beyond Eq.~(32) so that our results are applicable for the Type-I interactions 
considered here.}  For the singles spectral rates, we adapt the approach of Bennink \cite{Bennink}, who only considers collinear SPDC geometries.  Finally, we adapt the formalism 
of Mitchell \cite{Mitchell} to include spectral filtering.

Briefly, the procedure involves the quantum mechanical interaction Hamiltonian for the SPDC process, which is used to predict the transition rate for populating the 
initially empty signal and idler single-spatial modes (so-called `target' modes) with a biphoton state under the assumption of weak conversion so that perturbation 
theory applies \cite{MandelWolf}.   The interaction Hamiltonian involves an integral of the product of the pump, signal and idler modes over the volume of the nonlinear
 optical crystal and accounts for the phase mismatch between the beams.  We assume that the paraxial approximation holds for all beams, the pump beam is in a coherent state 
and that the transverse spatial profiles of the pump beam as well as modes collected by the optical fibers are given by a lowest-order Gaussian functions
\begin{equation}
U_{j}(\vec{r}) = e^{-(x_{j}^{2}+y_{j}^{2})/W_{j}^{2}}~(j=p,s,i),
\end{equation}
where $W_{j}$ is the 1/$e$ radius of the field mode.  Furthermore, we assume that the transverse extent of the crystal is large enough that it does not cause diffraction of 
these modes, the crystal is thin enough so that the Guoy phases of the Gaussian beams are constant over the length of the crystal, birefringent walk-off can be ignored for the
 thin crystal, and that the pump beam is monochromatic so that there is a definite relation between the signal and idler frequencies via the energy conservation relation.  

The fiber collection modes, assumed to lie in the $x-y$ plane and described using the primed coordinate systems shown in Fig.~\ref{pumpgeo}, are back-propagated from the fiber, 
through the imaging lenses, and to the nonlinear optical crystal, where they experience refraction at the air-crystal interface.  In general, the angle of refraction is frequency 
dependent due to the crystal dispersion, but we find the spectral rates are modified only slightly from this effect.  Therefore, for simplicity, we take the angle of refraction 
to be constant at the value determined from Snell's law using the crystal refractive index at the central frequencies of the signal and idler beams, resulting in target modes in 
the crystal described by the unprimed coordinate systems shown in Fig.~\ref{pumpgeo}, which can be transformed to the pump-beam coordinate system in terms of the signal (idler) 
emission angle $\theta_{s(i)}$  inside the crystal.  The expressions for the spectral rates involve a factor related to the geometrical overlap of the spatial mode envelopes 
(see the Eq. (21), Fig.~2, and the related discussion in Ling \textit{et al.} \cite{Ling}), reducing the rates when the beams do not overlap.  This effect is has only a minor 
impact ($<2\%$) on our predictions for the small emission angles considered here and for the tightest focusing conditions used in the simulations and experiments described below.
To support our goal of obtaining analytic predictions that can be interpreted readily, we ignore this effect.  

\subsection{Joint Spectral Rate}

Following this procedure and under the various assumptions described above, we find that the joint spectral rate is given by \cite{Ling} 
\begin{equation}
\frac{dR}{d\omega_{s}}=\frac{\eta_s \eta_i 
Pd_{\textrm{eff}}^2\alpha_{s}^2\alpha_{i}^2\alpha_{p}^2\omega_{s}\omega_{i}}{\pi \epsilon_{0}c^3 n_{s} n_{i} n_{p}}|\Phi(\Delta \vec{k})|^2,
\label{jointspectralrate}
\end{equation}
which is the rate of biphoton production per unit frequency interval of the signal beam.  Here,  $\eta_{s(i)}$ is the overall efficiency of the signal (idler) paths, 
including any losses due Frensel reflections, absorption, and the detector quantum efficiency, $d_{\textrm{eff}}$ is the effective second-order nonlinear coefficient 
for the crystal, $P$ is the average pump power, $\epsilon_{0}$ is the permitivity of free space, $c$ is the speed of light in vacuum, and $\alpha_{j} = \sqrt{2/(\pi W_j^2)}$, 
are the Gaussian mode normalization constants.  In Eq.~\ref{jointspectralrate}, the fact that the differential of $R$ with respect to frequency is in terms of the signal 
frequency $\omega_s$ reflects our assumption that the pump-beam frequency is monochromatic and hence $|d\omega_s|=|d\omega_i|$. 

The efficiency function in Eq.~\ref{jointspectralrate} is given by
\begin{equation}
\Phi(\Delta \vec{k}) = \frac{\pi L}{\sqrt{AC}}e^{-{\Delta k_{y}^{2}/(4C)}}\,\mathrm{sinc}\left(\frac{\Delta k_{z}L}{2}\right).
\label{phijoint}
\end{equation}
Here,
\begin{equation}
 A = \frac{1}{W_{p}^{2}}+ \frac{1}{W_{s}^{2}}+ \frac{1}{W_{i}^{2}},
\end{equation}
and
\begin{equation}
C = \frac{1}{W_{p}^{2}}+ \frac{\cos^{2}\theta_{s}}{W_{s}^{2}}+ \frac{\cos^{2}\theta_{i}}{W_{i}^{2}}.
\end{equation}
The phase mismatch is defined through the relations 
\begin{eqnarray}
\Delta \vec{k} & = & \vec{k}_{p}-\vec{k}_{s}-\vec{k}_{i} \\
& = & \Delta k_y \hat{y} + \Delta k_z \hat{z},
\end{eqnarray}
with $\vec{k}_{j} = n_{j}(\omega_j,\vec{k}_j)\omega_{j}\hat{k}_{j}/c $, where $n_{j}$ is the frequency- and angle-dependent 
refractive index, $\hat{k}_{j}$ is the propagation unit vector for mode $j$, and $\hat{x},\hat{y}$ are the unit vectors for the pump-beam coordinate system \cite{Beouff}. 
 We assume that the wavevectors lie in the $y-z$ plane and that $k_p$ is in a plane containing two of crystal axes for the case of the biaxial BiBO crystal considered here.  
See Fig.~\ref{pumpgeo}(b) for an illustration of the phase mismatch and its relation to the emission angles.  In terms of the wavevector magnitudes
\begin{eqnarray}
\Delta k_z & = & k_p - k_s \cos \theta_s - k_i \cos \theta_i \label{zphasemismatch} \\
\Delta k_y & = & k_s \sin \theta_s - k_i \sin \theta_i.
\label{yphasemismatch}
\end{eqnarray}

The total joint rate is found from the joint spectral rate and is given by
\begin{eqnarray}
R & = &\frac{\eta_s \eta_i 
Pd_{\textrm{eff}}^2\alpha_{s}^2\alpha_{i}^2\alpha_{p}^2\omega_{s}\omega_{i}}{\pi \epsilon_{0}c^3 n_{s} n_{i} n_{p}}  \nonumber \\
& \times & \int_{-\infty}^\infty T_s(\omega_s)T_i(\omega_p-\omega_s) |\Phi(\Delta \vec{k})|^2\,d\omega_s,
\label{totaljointrate}
\end{eqnarray}
where $T_{s(i)}$ is the intensity transmission function of spectral filters placed in the signal (idler) path.  

The phase mismatch (Eqs.~\ref{zphasemismatch} and \ref{yphasemismatch}) plays an important role in determining the rates, as can be seen in Eq.~\ref{phijoint}.  
We find that a physical explanation for the results presented below is more easily understood by considering a Taylor-series expansion of the wavevector magnitudes of 
the signal and idler beams with respect to frequency and truncating the series after second order.  Following this approach, we have
\begin{equation}
k_{s(i)}(\omega) \simeq k_{s(i)0}+\frac{n_{g,s(i)}}{c}(\omega-\omega_{s(i)0})+\frac{1}{2}k''_{s(i)}(\omega-\omega_{s(i)0})^2,
\end{equation}
where $k_{s(i)0}=k|_{\omega=\omega_{s(i)0}}$ is the wavevector magnitude, $n_{g,s(i)}=c\, \partial k_{s(i)}/\partial \omega_{s(i)}|_{\omega=\omega_{s(i)0}}$ is the group index, 
and $k''_{s(i)}= (1/c)(\partial n_{g,s(i)}/\partial \omega_{s(i)}|_{\omega=\omega_{s(i)0}})$ is the group velocity dispersion parameter, all evaluated at the carrier frequency
 $\omega_{s(i)}$ of the signal (idler) beam.  The carrier frequencies are chosen so that the interaction is perfectly phased matched ($\Delta \vec{k}=0$) for these
 frequencies. The target modes are chosen by adjusting the angle between the pump wave vector and the crystal axes, $\theta_p$, so that
\begin{eqnarray}
k_p - k_{s0} \cos \theta_s - k_{i0} \sin \theta_i & = &0 \\
k_{s0} \cos \theta_s - k_{i0} \cos \theta_i & = & 0.
\end{eqnarray}

For the case of frequency-degenerate down-conversion ($\omega_{s0}=\omega_{i0}=\omega_p/2$), the angles of emission are identical ($\theta_s=\theta_i \equiv \theta$),
 as well as the group indices ($n_{g,s}=n_{g,i} \equiv n_g$) and group velocity dispersion parameters ($k''_s=k''_i \equiv k''$) resulting in the relations
\begin{eqnarray}
\Delta k_z & = & - k'' \cos \theta \, (\omega_s-\omega_p/2)^2 \label{approxkz} \\
\Delta k_y & = & 2 n_g \sin \theta (\omega_s-\omega_p/2)/c. \label{approxky}
\end{eqnarray}
Hence, the longitudinal phase mismatch (the $z$-component) is quadratic in frequency\footnote{The error in Eq.~(32) in Ling \textit{et al.} \cite{Ling} stems from the fact 
that they assumed a linear relation between $\Delta k_z$ and frequency, which is not valid for frequency-degenerate Type-I SPDC as shown here.} and the transverse phase mismatch 
(the $y$-component) is linear in frequency, vanishing for collinear down-conversion ($\theta=0$).  For nondegenerate down-conversion, both $\Delta k_z$ and $\Delta k_y$ are 
dominantly linear functions of frequency.  These observations have important implications for the spectral bandwidth of the down-converted light as discussed in Sec.~III.

\subsection{Singles Spectral Rate}
We calculate the singles spectral rate in a similar manner as the joint spectral rate.  Briefly, the singles spectral rate for the signal (idler) is given by the joint spectral rate, but for emission into \textit{any} emission direction for the idler (signal).   This is accomplished formally by defining a generalized efficiency function (see Eq.~\ref{phijoint}) that determines the overlap of the pump and signal (idler) modes and the entire set of transverse modes of the idler (signal)  \cite{Bennink,Uren2}.  For this procedure, we use the complete set of orthonormal Hermite-Gauss modes with generalized beam normalization parameters for the signal beam given by
\begin{equation}
\alpha_{s}^{(n,m)}= \sqrt{\frac{2}{2^{n+m}n!m!\pi W_{s}^2}} \,;
\end{equation}
a similar definition for the idler beam parameter is obtain by substitution $s \rightarrow i$.

We find that the singles spectral rate for the signal beam is given by
\begin{equation}
\frac{dR_s}{d\omega_{s}}=\sum_{n,m=0}^{\infty} \frac{\eta_s \eta_i 
P d_{\textrm{eff}}^2 (\alpha_s^{(n,m)})^2 \alpha_{i}^2\alpha_{p}^2\omega_{s}\omega_{i}}{\pi \epsilon_{0}c^3 n_{s} n_{i} n_{p}} |\Phi_s^{(n,m)}(\Delta \vec{k})|^2.
\label{singlesspectralrate}
\end{equation}
The generalized mode efficiency functions $\Phi_s^{(n,m)}$ can be determined analytically, although the expressions are lengthy for large $(n,m)$.  We find that all odd numbered modes in the $x$ direction ($n$ odd) are zero due to the fact that $\Delta k_{x} = 0$.  It is instructive to analyze the lower-order mode contributions to develop physical intuition underlying the differences between the joint and singles spectral rates.  In particular, we find that 
\begin{align}
 \Phi_s^{(0,1)}(\Delta \vec{k})&= \frac{ i\pi \sqrt{2}e^{-\Delta k_{y}^{2}/(4C)}}{W_{i}A^{3/2}\sqrt{C}} \nonumber \\
 & \times \bigg\{\cos \theta_{i}\Delta k_{y}L\,\mathrm{sinc}\left[\frac{\Delta k_{z} L}{2}\right] \nonumber \\
 &+(\cos \theta_{i}D+\sin \theta_{i}2C) \nonumber \\
& \times \frac{L}{\Delta k_{z}}\left(\cos\left[\frac{\Delta k_{z}L}{2}\right]-2\,\mathrm{sinc}\left[\frac{\Delta k_{z}L}{2}\right]\right)\bigg\},
\label{phi01}
\end{align}
and
\begin{align}
 \Phi_s^{(2,0)}(\Delta \vec{k})&= \frac{2\pi}{\sqrt{AC}}\left(\frac{2}{AW_{i}^{2}}-1\right)\nonumber\\
 &\quad \times e^{-\Delta k_{y}^{2}/(4C)}L{\rm sinc}\left[\frac{\Delta k_{z} L}{2}\right],
\label{phi20}
\end{align}
where
\begin{equation}
D = \frac{\rm{sin}(2\theta_{s})}{W_{s}^{2}}-\frac{\rm{sin}(2\theta_{i})}{W_{i}^{2}}.
\end{equation}

Finally, the total singles rate for the signal is given by
\begin{eqnarray}
R_s & = & \sum_{n,m=0}^{\infty}\frac{\eta_s 
Pd_{\textrm{eff}}^2(\alpha_{s}^{(n,m)})^2\alpha_{i}^2\alpha_{p}^2\omega_{s}\omega_{i}}{\pi \epsilon_{0}c^3 n_{s} n_{i} n_{p}}  \nonumber \\
& \times & \int_{-\infty}^\infty T_s(\omega_s)|\Phi_s^{(n,m)}(\Delta \vec{k})|^2\,d\omega_s,
\label{totalsinglesrate}
\end{eqnarray}
Similar expressions can be obtained for the idler spectral and total rates by appropriately substituting $s \rightarrow i$.  We find good convergence in our predictions when we truncate the sums in Eqs.~\ref{singlesspectralrate} and \ref{totalsinglesrate} above $n$ and $m$ above 10.

\section{Spectral Rates for Different Configurations}

In this section, we use the formalism developed above to predict the spectral rates for four different configurations.  The base configurations for all results assume the use of a 
600-$\mu$m-long BiBO crystal pumped with a modelocked 355-nm-wavelength laser with a $\sim$10-ps-long pulse duration and an average power of $P=1$ mW.  The spectral width of the pump
 light is 0.013 nm (sech$^2$ pulse shape), which is much less than the spectral width of the down-converted light and hence the formalism developed above, which assumes a monochromatic
 pump beam, is applicable \cite{Castelletto}.  We consider two different focusing conditions:  ``loose'' focusing with $W_p=250$ $\mu$m and $W_s=W_i=100$ $\mu$m and ``tight'' focusing
 with $W_p=150$ $\mu$m and $W_s=W_i=50$ $\mu$m.  We also consider the ``degenerate'' SPDC configuration where $\omega_{s0}=\omega_{i0}$ (corresponding to a central wavelength of
 710 nm) and ``nondegenerate'' configuration where the signal (idler) frequency corresponds to a wavelength of 850 nm (609.6 nm).  We take the signal and idler transmission/detection
 losses to be zero so that $\eta_s=\eta_i=1$.

\begin{figure}
\begin{center}
 \includegraphics[scale=.45]{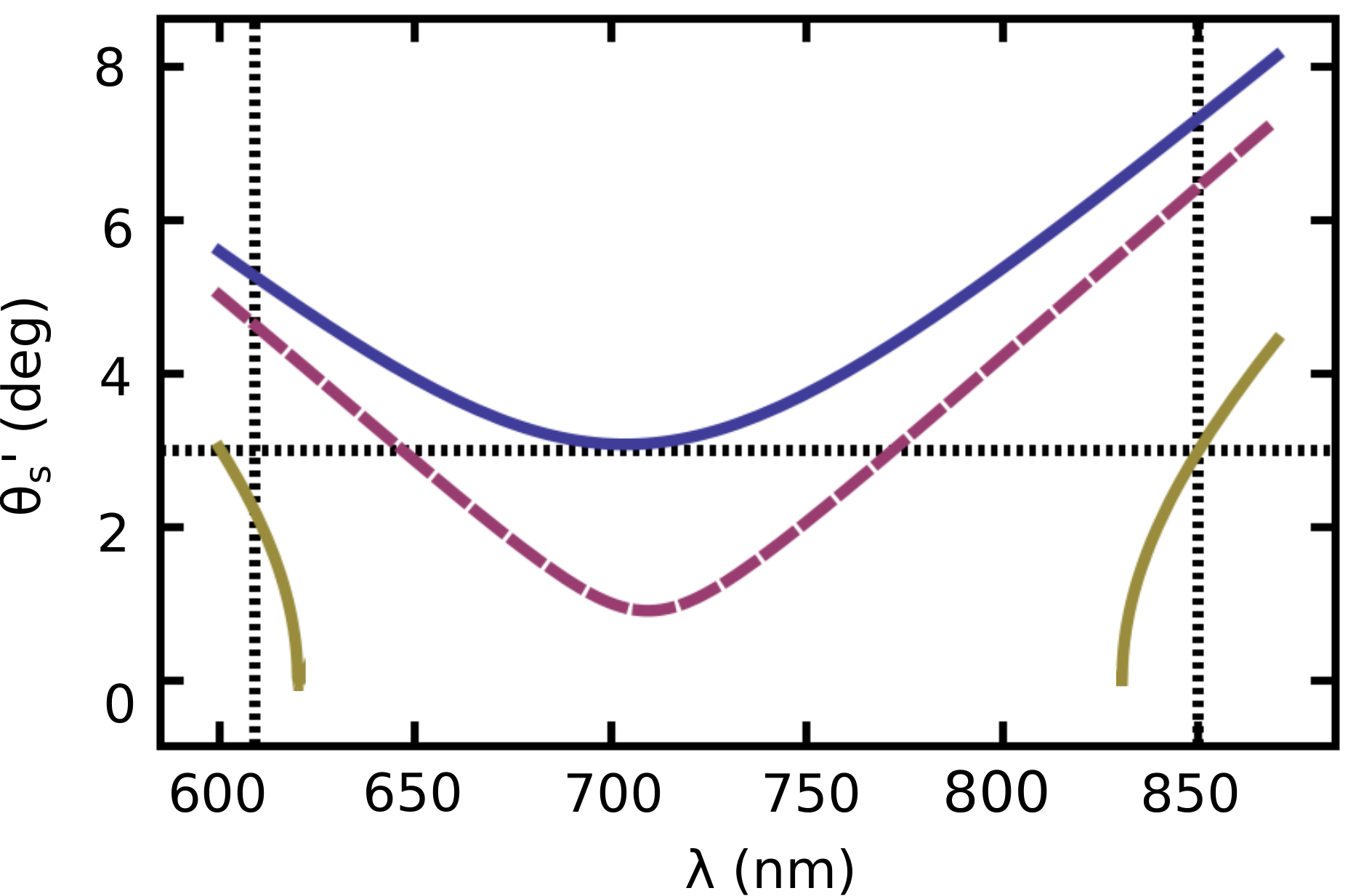}
\caption{\textbf{Opening angle versus wavelength} Exterior opening angle as a function of wavelength for crystal tilt angles: $\theta_{p}=141.9^{\circ}$ (blue line),
 $\theta_{p}=142.2^{\circ}$ (maroon dashed line), and $\theta_{p}=143.0^{\circ}$ (gold line). The curves intersect this line in two places showing that both wavelengths will 
be detected in the fiber. The vertical dashed lines represent a conjugate pair of wavelengths. These show that the two angles in the same direction are similar to, 
but not conjugate wavelengths for noncollinear geometries.}
\label{anglevswave}
\end{center}
\end{figure}

BiBO is a biaxial crystal that can be phase matched in our spectral range for the case when the pump beam experiences an angle-independent refractive index and the down-converted 
light experiences an angle-dependent refractive index.  The nonlinear coefficient $d_\mathrm{eff}$ for BiBO is larger than the commonly used uniaxial crystal BBO when the pump beam 
makes an azimuthal angle of 90$^\circ$ with respect to the BiBO crystal axes and the polar angle is adjusted to obtain the desired SPDC geometry \cite{Hellwig, Ghotbi}. 
 Figure~\ref{anglevswave} shows the angle of emission for the signal beam outside the crystal $\theta'_s$ that gives rise to perfect phase matching ($\Delta \vec{k}=0$) as a 
function of wavelength for BiBO and for different phase matching angles $\theta_p$.  For nondegenerate SPDC, there are two solutions to the phase matching condition so that two 
spectral peaks are expected in both the signal and idler paths.  These do not necessarily give rise to correlated photons and hence some care is needed to make sure that these spectral
 bands are separated and subsequently filtered.

\subsection{Degenerate Collinear SPDC}

\begin{figure}[t]
\begin{center}
 \includegraphics[scale=.38]{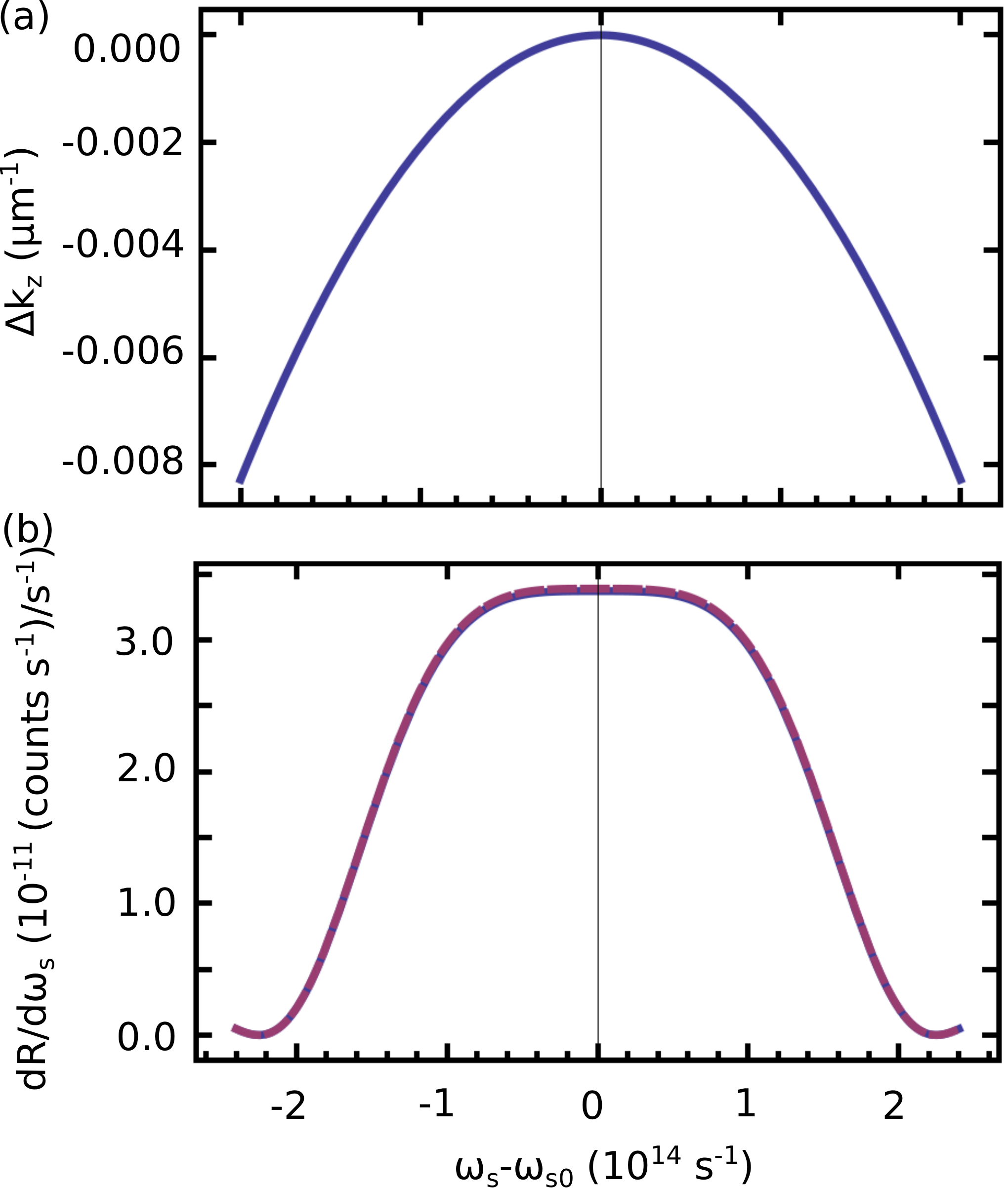}
\caption{\textbf{Collinear Degenerate} (a) The longitudinal phase mismatch as a function of frequency ($\omega_{s0} = 2.65\times10^{15}$ $s^{-1}$) for the collinear degenerate case. (b) The joint spectral rate (purple, dashed) and the singles spectral rate (blue, solid) are essentially equal.  Here, $\theta_p=142^\circ$.}
\label{COLD}
\end{center}
\end{figure}

Here, we consider the case when the down-converted light is emitted in the same direction as the pump beams ($\theta_s=\theta_i=0$) and at the same frequencies. 
 This configuration is not practical because it is not possible to separate the signal and idler beams, but it illustrates some important physics of the interaction. 
 Figure~\ref{COLD}(a) shows the longitudinal phase mismatch as a function of frequency, where it is clearly seen that it is quadratic in frequency as predicted by Eq.~\ref{approxkz}. 
 Furthermore, as seen in Eq.~\ref{approxky}, $\Delta k_y \sim 0$ (not shown).  In this case, the exponential term in $\Phi(\Delta \vec{k})$ (Eq.~\ref{phijoint}) approaches unity
 and the efficiency is dominated by the sinc function; hence, the crystal length dictates the joint spectral rate shown in Fig.~\ref{COLD}(b) for the ``loose'' focusing condition.   
In particular, the width of the joint spectral rate can be determined approximately from the condition $\Delta k_z L \sim 2\pi$ and is independent of the focusing conditions.  
Given the quadratic nature of $\Delta k_z$, the joint spectral rate is quite broad and nearly constant about the central frequency.

For the collinear case considered here, the higher-order spatial mode contributions ($n,m > 0$) to the singles spectral rates are very small.  In particular, $D=0$ so that $\Phi_s^{(0,1)} \sim 0$ (see Eq.~\ref{phi01}).  The frequency-dependent part of $\Phi_s^{(2,0)}$ arises from the sinc function and has an identical form as $\Phi$ so that the singles and joint spectral rates should have an identical frequency dependence.  Furthermore, the scale factor in Eq.~\ref{phi20} can be made small using appropriate focusing.  In particular
\begin{equation}
\left\lvert \frac{2}{A W_s^2}-1 \right\rvert=\frac{1}{1+2 W_p^2/W_s^2} \ll 1
\label{focusing}
\end{equation}
when $W_p \gg W_i$, where we have assumed that $W_s=W_i$, which is known to maximize $R$ \cite{Ling}.  Hence, the higher-order-mode efficiency functions can be made small
 for the case when the pump is focused more loosely than the signal and idler modes.  In this case, the singles spectral rate should be nearly identical to the joint spectral 
rate, as indeed is supported by the data shown in Fig.~\ref{COLD}(b).  Thus, heralding efficiency could be very high (if the signal and idler beams could be spatially separated)
 because the spectral rates are essentially the same.

The focusing condition for high heralding efficiency is opposed to the case for maximizing the total joint rate, which is optimized when the pump beam is focused tighter than
 the signal and idler modes \cite{Bennink,Ling,Dixon}.  However, as we will see below, the count rate can be increased while maintaining or even improving the heralding efficiency
 by focusing all beams to a smaller waist but keeping the pump beam waist larger than the signal and idler waists.

\subsection{Degenerate Noncollinear SPDC}

To spatially separate the signal and idler beams for the case of degenerate SPDC, the crystal must be tilted by a small amount to achieve noncollinear phase matching, which 
results in the down converted light being emitted in a cone surrounding the pump beam with a small opening angle, typically 3$^\circ$ outside the crystal in a typical experiment. 
 The correlated signal and idler beams are then located on opposite sides of the cone, as indicated in Fig.~\ref{pumpgeo}(a).

\begin{figure}
\begin{center}
 \includegraphics[scale=.38]{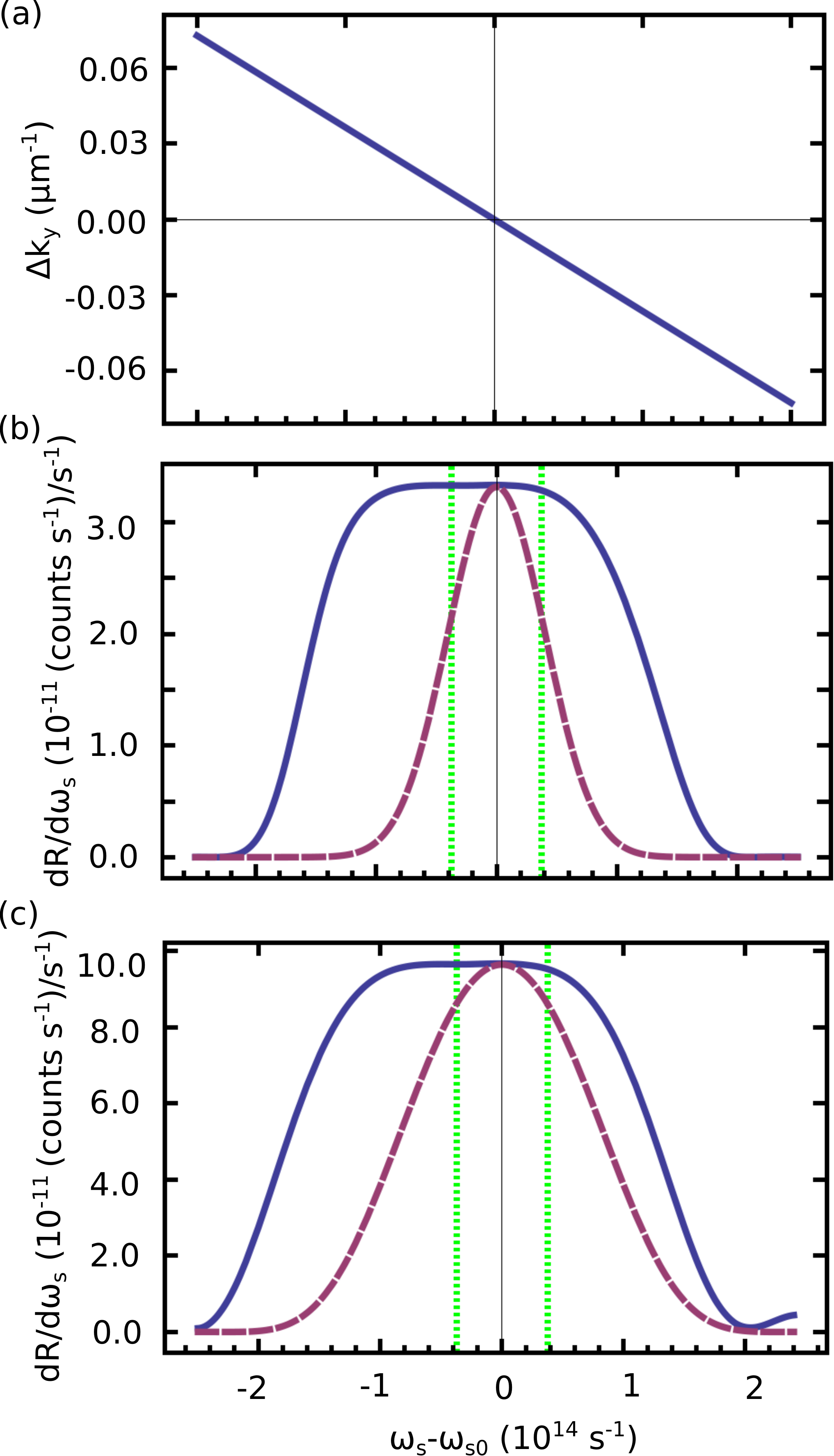}
\caption{\textbf{Noncollinear Degenerate} (a) The transverse phase mismatch as a function of frequency for $\omega_{s0}=2.65\times10^{15}$ $s^{-1}$ for the  
noncollinear degenerate case. Joint spectral rate (purple, dashed) and singles spectral rate (blue, solid) for (b) ``loose'' and (c) ``tight'' focusing conditions.  
The green  vertical lines at frequency offsets of $\pm 3.73 \times 10^{13} s^{-1}$  correspond to a full bandwidth of $\sim$20 nm for a central wavelength of 710 nm. Here, 
$\theta_p=141.9^\circ$, $\theta_s=\theta_i=1.64^\circ$, corresponding to $\theta'_s=\theta'_i=3.04^\circ$.}
\label{NONCOLD}
\end{center}
\end{figure}

For degenerate noncollinear down conversion, $\Delta k_z$ is essentially identical to that shown in Fig.~\ref{COLD}(a) (see Eq.~\ref{approxkz}), but $\Delta k_y$ becomes large and 
takes on a nearly linear frequency dependence as seen in Fig.~\ref{NONCOLD}(a) and predicted by Eq.~\ref{approxky}.  For either of focusing conditions, the joint spectral rate is 
dominated by the transverse phase mismatch $\Delta k_y$ and the width of the joint spectral is given approximately by $\Delta k_y \sim 2 \sqrt{C}$.  The parameter $C$ is inversely 
related to the beam waists and hence we expect that the joint spectral rate is now strongly dependent on the focusing conditions, being broader for tighter focusing, and largely 
independent of the crystal length.  Figure ~\ref{NONCOLD}(b) and (c) show the joint spectral rate for the ``loose'' and ``tight'' focusing conditions, respectively, where tighter focus broadens the spectrum, as expected.  Furthermore, because of the linear frequency dependence of $\Delta k_y$, the spectral shape is nearly Gaussian due to the 
exponential term in Eq.~\ref{phijoint}.
The singles spectral rates are substantially different in comparison to the joint spectral rate for the noncollinear case, as seen in Figs.~\ref{NONCOLD}(b) and (c).  
This arises from the fact that a photon detected in the single mode of the signal path has its corresponding photon emitted into a higher-order mode in the idler path. 
 In particular, $\Phi_s^{(0,1)}$ does not vanish and the term shown in the second line of Eq.~\ref{phi01} is linear in $\Delta k_y$, effectively broadening the spectrum. 
 Indeed, in summing over all modes, we see in Fig.~\ref{NONCOLD}(b) that the singles spectral rate is much broader than the joint spectral rate and, by comparing panels 
(b) and (c), there is only a weak dependence in the singles spectral width for different focusing conditions.  Inspection of panels (b) and (c) also reveals that tighter 
focusing gives rise to a substantially higher biphoton generation rate with the tighter focusing condition giving nearly a factor of 3 increase in the rate in comparison to the
 loose focusing condition.

Based on this data, we anticipate that the heralding efficiency will be quite low for this noncollinear geometry, which will be explored quantitatively in Sec. IV. 
 It is also clear why spectral filter can help improve the heralding efficiency.  For a narrow-band spectral filter centered on the degenerate frequency, the integral 
of the joint spectral rate and the singles spectral rates (needed to determine $R$ and $R_{s(i)}$) over the filter bandwidth (vertical lines) can be made similar to each other. 
 Furthermore, tighter focusing broadens the joint spectral rate while leaving the singles spectral rate approximately the same, so we expect higher heralding efficiency for this
 focusing condition.

\subsection{Collinear Nondegenerate SPDC}

\begin{figure}
\begin{center}
 \includegraphics[scale=.38]{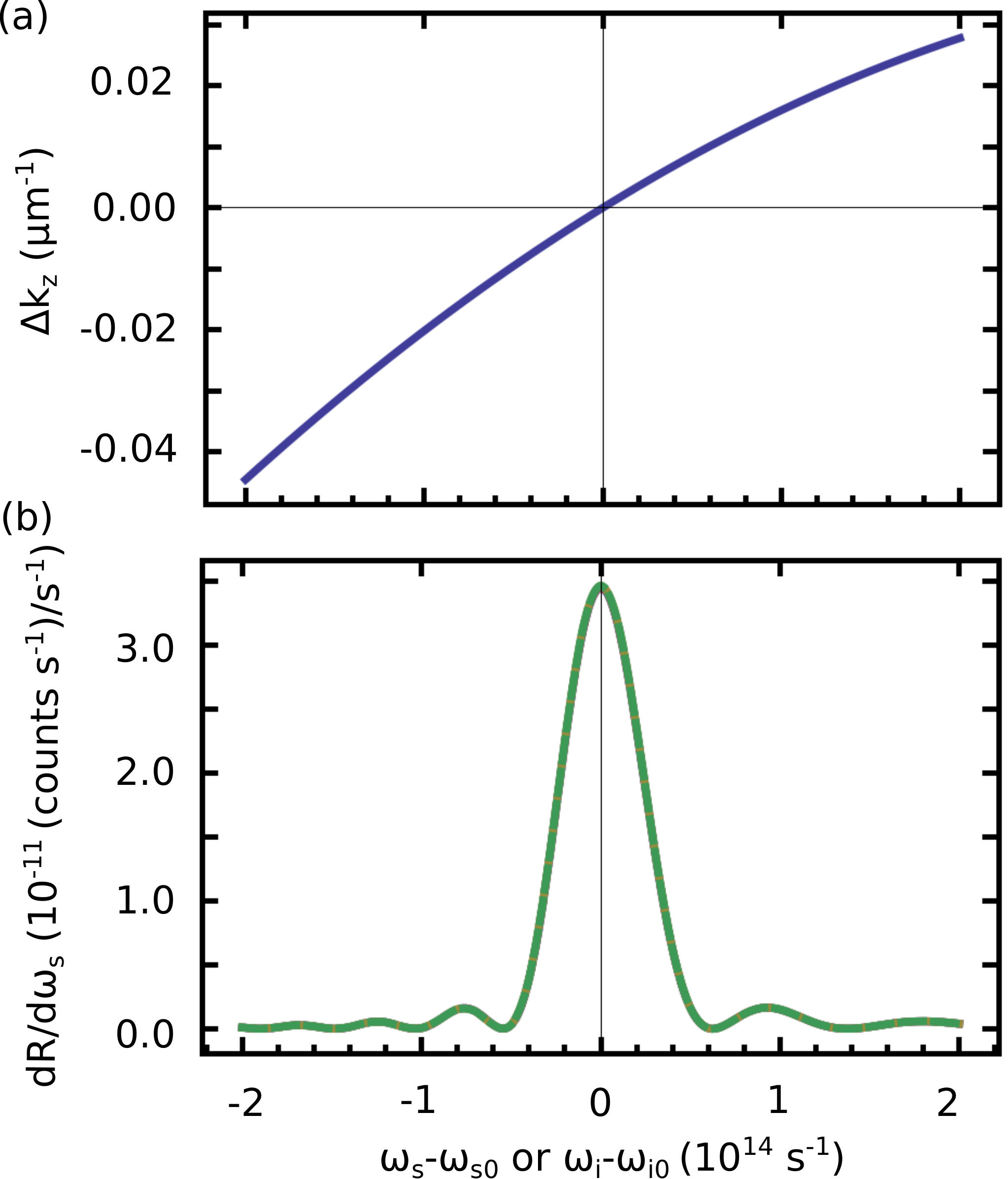}
\caption{\textbf{Collinear Nondegenerate} (a) Longitudinal phase mismatch as a function of frequency. (b) The joint spectral rate (purple, dashed) and the singles spectral rates (green, solid) are essentially identical; there differences cannot be discerned.  Here, $\omega_{s0}=3.09\times10^{15}$ $s^{-1}$ (corresponding to a wavelength of 850 nm), $\omega_{i0}=2.22\times10^{15}$ $s^{-1}$ (corresponding to a wavelength of 609.6 nm), and $\theta_p=143.22^\circ$.}
\label{COLND}
\end{center}
\end{figure}

We now consider the case of tilting the crystal to obtain collinear down conversion with different signal and idler frequencies.   
 As in the collinear degenerate case, $\Delta k_y \simeq 0$, and the longitudinal phase mismatch is a nearly linear function of frequency, as seen in Fig.~\ref{COLND}(a), 
and hence the joint spectral rate has a width that is substantially narrower as seen in Fig.~\ref{COLND}(b).  As in the degenerate collinear case, the shape of the spectrum 
is dictated by the sinc function in Eq.~\ref{phijoint} and hence depends mostly on the crystal length and is independent of the focusing conditions.  Also, for the case when 
the pump waist is much larger than the target-mode waists, the singles spectral rate is essentially identical to the joint spectral rate (see Fig.~\ref{COLND}) for the same 
reasons as discussed above for the degenerate-frequency case.  With this observation, we expect high heralding efficiency without the use of any spectral filtering.  
While not shown here, the magnitude of the spectral rates all increase for tighter focusing.

\subsection{Noncollinear Nondegenerate SPDC}

\begin{figure}
\begin{center}
 \includegraphics[scale=.38]{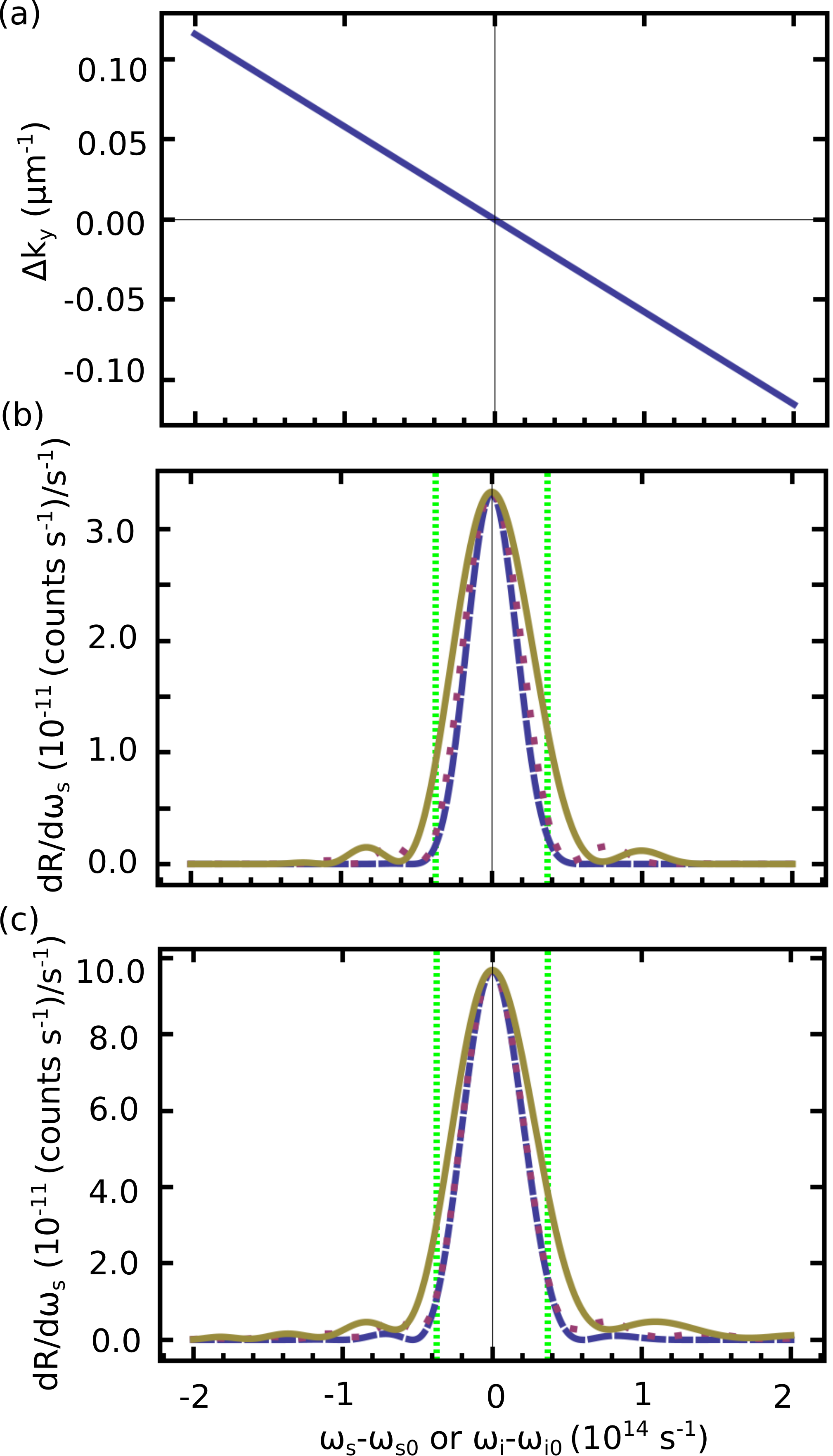}
\caption{\textbf{Noncollinear Nondegenerate} (a) Transverse phase mismatch as a function of frequency for the noncollinear nondegenerate case.
 The joint spectral rate  (blue, dashed) and the singles spectral rate for the signal (gold, solid) and idler (purple, dotted) for the (b) ``loose'' and 
(c) ``tight'' focusing conditions.  The carrier frequencies are the same as those given in the caption to Fig.~\ref{COLND}, 
$\theta_p=142.44^\circ$, $\theta_s=3.05^\circ$, $\theta'_s=5.62^\circ$, $\theta_i=2.17^\circ$, and $\theta'_i=4.02^\circ$.}
\label{NONCOLND}
\end{center}
\end{figure}

For the noncollinear situation, both the longitudinal and transverse phase mismatch play a role in determining the spectral rates.  
The longitudinal mismatch is essentially identical to the collinear case shown in Fig.~\ref{COLND}(a) for the small emission angles considered here, 
and the transverse phase mismatch is large and an approximately linear function of frequency as seen in Fig.~\ref{NONCOLND}.  In this situation, the 
longitudinal phase mismatch is somewhat dominant so that the width of the joint spectral rate is only weakly dependent on the focusing conditions, 
as seen in Fig.~\ref{NONCOLND}(b) and (c), with a slight broadening for tighter focusing.  Interestingly, the singles spectral rate is only slightly 
broader than the joint spectral rate so that the heralding efficiency should be higher in this case in comparison to the noncollinear degenerate case. 
 Spectral filtering will also increase the heralding efficiency, although a much narrower filter is needed to accomplish this task given that the overall 
spectrum is narrower than for the degenerate case.  As before, the overall rate is increased using the tighter focusing conditions.

\section{Heralding Efficiency and Joint Count Rates}
As mentioned above, high heralding efficiency is obtained when the joint spectral rate is similar to the singles spectral rates, which can be forced to be more similar using 
filters centered on the signal and idler carrier frequencies.   The ideal filter has a flat top profile of full width $\Delta \omega_f$ and unit transmission in the passband, 
which we assume in the plots in this section.   We do not consider the collinear degenerate case further because of the difficulty in spatially separating the signal and idler beams.

\begin{figure}
\begin{center}
 \includegraphics[scale=.38]{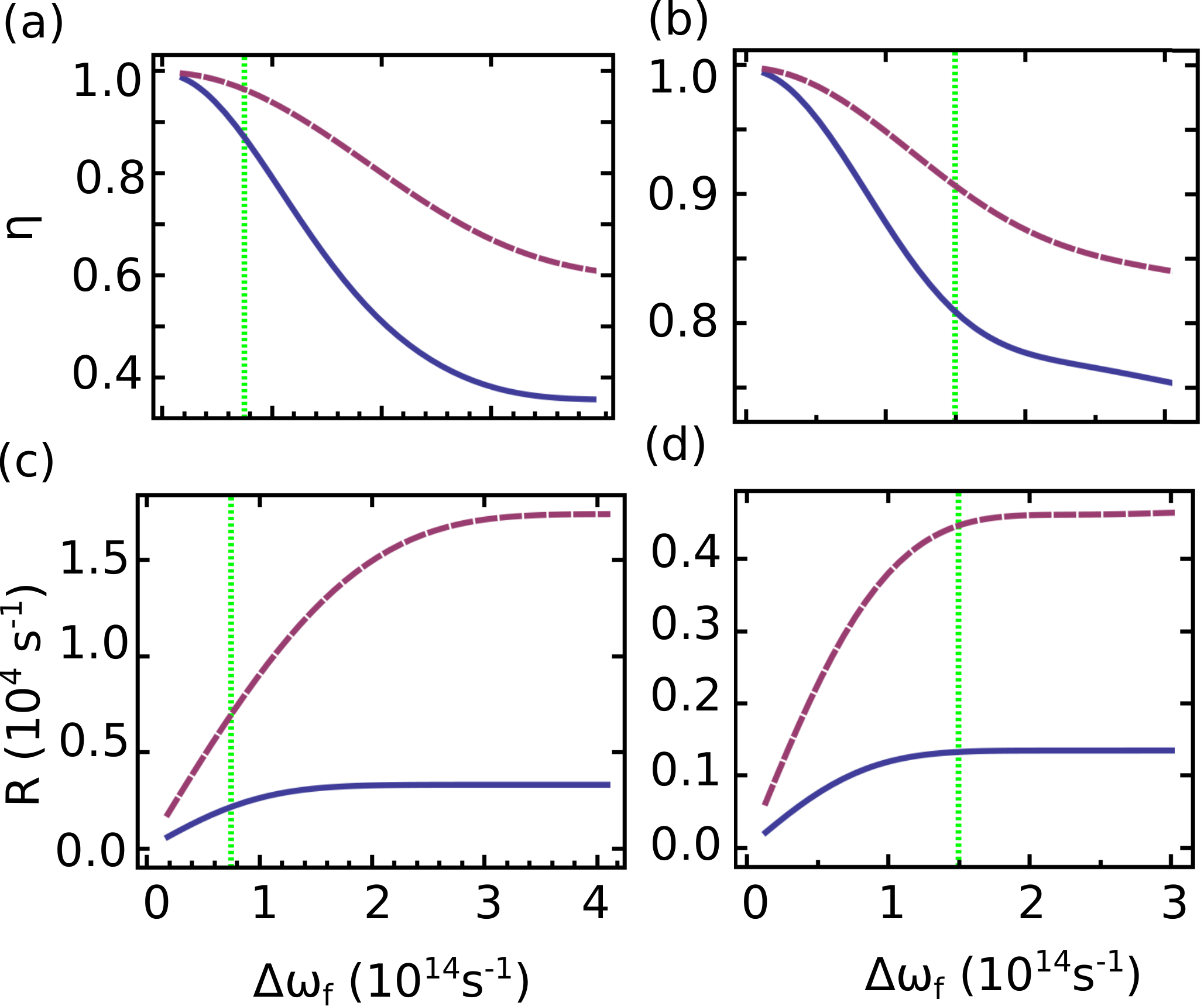}
 \caption{\textbf{Heralding efficiency and joint counts versus filter bandwidth.} Purple (dashed) lines have waist parameters in the crystal of 
$W_{p} = 150 \mu m$, $W_{s} = W_{i} = 50 \mu \textrm{m}$ and blue (solid) lines have waist parameters of 
$W_{p} = 300 \hspace{2pt}\mu  \textrm{m}$, $W_{s} = W_{i} = 100 \hspace{2pt}\mu \textrm{m}$. 
Vertical green lines at $7.46 \times 10^{13} s^{-1}$ show approximately where a 20 nm filter at 710 nm would be. (a) and (b) show the heralding efficiency as a function 
of filter frequency for the two sets of waist parameters for (a) noncollinear degenerate and (b) noncollinear nondegenerate. (c) and (d) 
show the joint count rates for (c) noncollinear degenerate and (d) noncollinear nondegenerate.}
\label{HE} 
\end{center}
\end{figure}

Figure~\ref{HE}(a) and (c) show the heralding efficiency and total joint rate for noncollinear degenerate SPDC as a function of filter bandwidth.  The vertical line indicates a 
filter with a $\sim$20 nm bandwidth.  It is seen that $\eta$ is quite low (below 60\%) for no spectral filtering (large $\Delta \omega_f$), but that the ``tight'' focusing condition
 has a substantially higher efficiency and rate for all filter bandwidths.  Heralding efficiencies $>90\%$ can be obtained for the tighter focusing and a narrow filter bandwidth, 
but at the cost of the total joint rate. 

For collinear nondegenerate SPDC, the heralding efficiency exceeds 99.5\% for both focusing conditions and without any spectral filtering.  
This is due to the similarity of the joint and singles spectral rates (Fig.~\ref{COLND}).  For ``loose'' (``tight'') focusing, $R=19.5$ kHz/s/mW ($R=56.8$ kHz/s/mW).
 Hence, this configuration with tight focusing is very promising for obtaining high efficiency and rates if high 
transmission filters can be identified to block the copropagating pump light and for spatially separating the signal and idler beams.

Noncollinear nondegenerate SPDC relaxes the constraints on the filters, but come at the price of a reduced $\eta$.  As for the degenerate case,
 $\eta$ is higher for tighter focusing (see Fig.~\ref{HE}(b), but the overall efficiency is much higher than the degenerate case.  
Because the joint spectral rate is much narrower, a narrower band filter is needed to obtain the highest efficiencies, but $\eta>90\%$ 
can be obtained with a 20-nm-bandwidth filter with almost no loss in total rate (see Fig.~\ref{HE}(d)).  Unfortunately, $R$ is substantially 
lower for the noncollinear nondegenerate rate because of the narrow spectrum in comparison to the noncollinear degenerate case with aggressive 
filtering, so the two geometries are nearly equivalent with respect to the efficiency and rate metrics.

\section{Comparison to Experimental Findings}

We compare our theoretical predictions to our experimental findings for the noncollinear degenerate case as well as the noncollinear nondegenerate case for a limited set of 
parameters. The pump beam is generated by a high-power modelocked laser (Coherent Palidan, 4 W maximum average power, 120 MHz pulse repetition rate, 10-ps-long pulse duration) 
focused into a 600-$\mu$m-long BiBO crystal (Newlight Photonics) anti-reflection coated at both the pump wavelength and the degenerate down conversion wavelength (710 nm).   
The down converted light is collected using a well-corrected achromatic lens (Schaefter and Kirchhoff GmbH, 60FC-T-0-M20l-02) and coupled into a single-mode fiber (Thorlabs custom fiber),
 that does not have anti-reflection coatings on the fiber ends for nondegenerate frequencies or anti-reflection coated fiber for degenerate frequencies (Oz Optics, custom fiber).  To coarsely align the signal and idler paths, we back-propagate laser light through the fibers and 
lenses toward the BiBO crystal at wavelengths close to the emitted SPDC light, adjusting the spot size to the desired values and placing the waist in the crystal, a process that 
is enabled using a beam profiler (Thorlabs, BP209-VIS).  Single photons are detected using silicon avalanche photodiodes operating in Geiger mode 
(Perkin-Elmer/Excelitas, SPCM - AQRH) with a peak quantum efficiency at 710 nm of $\sim 63\%$. The electrical pulses generated by the detectors are sent to a custom-programmed
 field-programmable gate array for coincidence counting ($\sim$9 ns coincidence window) and counting of single photon events.  To measure the singles spectral rate, we couple the 
light from the fibers into a triple monochrometer (Newport, Cornerstone 260 1/4 m) and a photomultiplier tube (Hammamatsu, H6780) followed by transimpedance amplification and 
lock-in detection.  The spectral response of the spectrometer system is calibrated using a high-pressure tungsten halogen lamp (Ocean Optics, HL2000).  

\begin{figure}
\begin{center}
 \includegraphics[scale=.4]{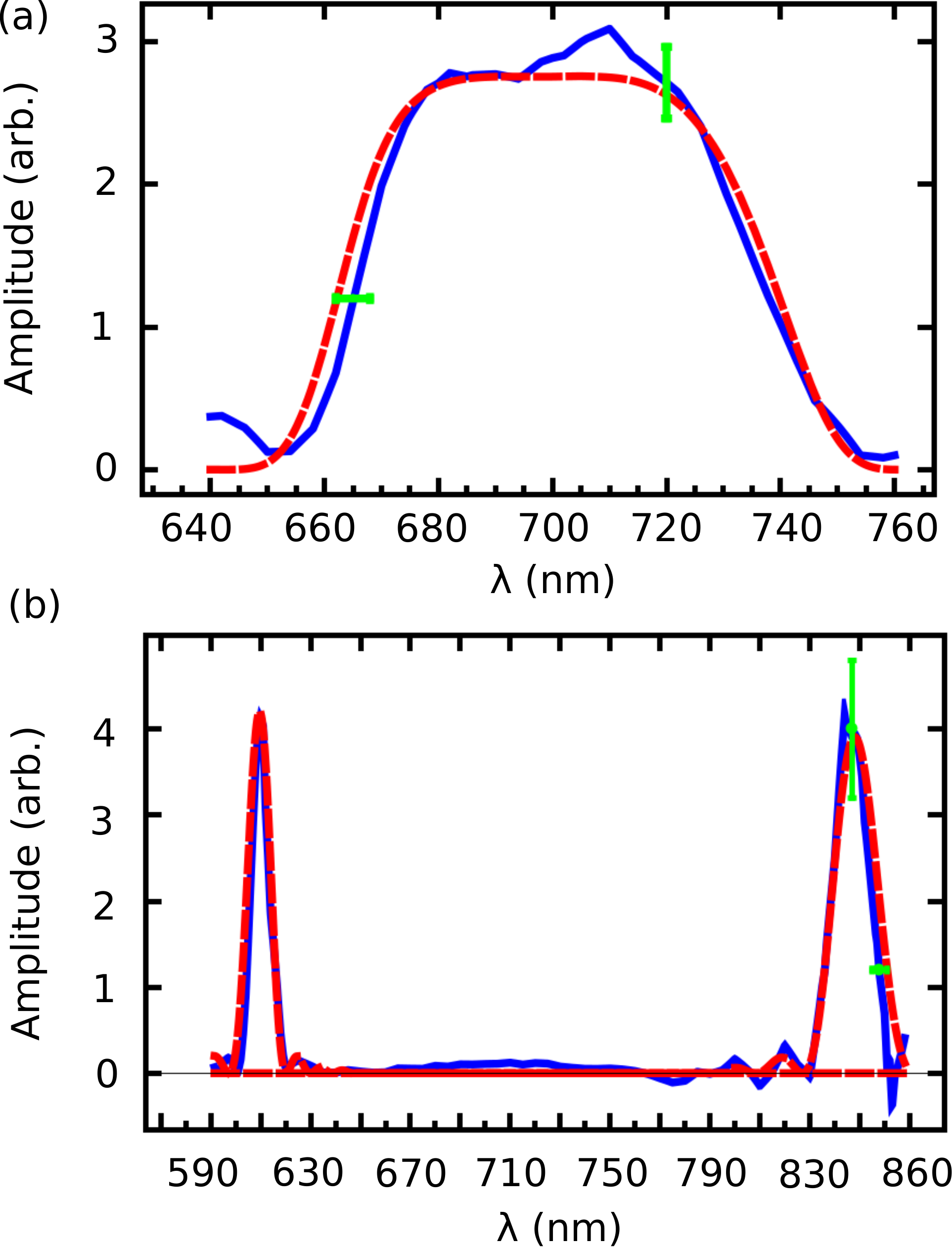}
\caption{\textbf{Predicted and measured spectra} Signal singles spectral rate for (a) noncollinear degenerate SPDC with $\theta'_s=3.04^{\circ}$ and (b) collinear 
nondegenerate SPDC.  Here, the experimental data is shown by the blue solid lines and the predictions are shown by the red dashed lines, 
where we use a least-square fit to choose the vertical scaling of the theoretical predictions.  The green bars show representative errors.}
\label{data}
\end{center}
\end{figure}

\textit{Noncollinear Degenerate} -- In light of the fact that the heralding efficiency is strongly dependent on the spectral rates, we measure the singles spectral rate for the 
signal beam, shown in Fig.~\ref{data}(a).  It is seen that the emission is very broad band and is in excellent agreement with our predictions.  Based on the discussion surrounding 
Fig.~\ref{NONCOLD}, spectral filter is required to obtain high heralding efficiency.  We use a 23-nm-bandwidth filter (Semrock TBP 704/13) that is close to an ideal top hat with an
 efficiency of $~99\%$, which we angle tune to center the passband on the degenerate wavelength.

Using the photon-counting setup, we measure $\eta=43\pm0.5\%$ with accidental coincidences contributing 0.5 $\%$ to this value 
(this value is not subtracted in the quoted efficiency) with $W_{p} = 250\pm5$ $\mu$m, $W_{s}= W_{i} = 100\pm5$ $\mu$m.  To compare to our theoretical predictions, 
we need to correct for the transmission/detection losses, which we estimate as $\eta_s=\eta_i=75.2\pm1.2\%$), with contributions of $62.5\pm0.5\%$ for the detection efficiency, 
$84\pm2\%$ for the non-ideal behavior of the spectral filters and minimal loss on the AR-coated fibers of $<1\%$. Using these values, we obtain a
 corrected heralding efficiency $\eta_{correct}=81.7\pm2.6\%$ , which should be compared to our theoretical prediction of $82.1 \pm 1.3 \%$.  The agreement is within our 
assigned errors and is excellent.  

To measure the total joint count rate, we adjust the average pump power to $\sim$100 mW so that the single count rates are well below the saturation rate of the detector and
 much larger than the dark count rate.  We then scale the results to determine the rate for $P=1$ mW so we can directly compare to our predictions.  We find that $R=982\pm20$ Hz.  
To find a corrected rate, we divide this result by the product $\eta_s \eta_i$ to arrive at $R_{correct}=2.39\pm0.05$ kHz, while we predict $R=2.16\pm0.13$ kHZ.  The agreement is 
within approximately a standard deviation of our measurements, which is good.

\textit{Noncollinear Nondegenerate} -- For the case of noncollinear nondegenerate SPDC with the signal (idler) central wavelength of 850 nm (609.6 nm) with the same waists
as above we measure $\eta=29.4\pm0.6\%$,
  with accidental coincidences contributing 0.4$\%$.  We do not use narrowband spectral filters for this measurement, but use a low-pass 
filter (Semrock FF01-650/SP-25, $>99.0\%$ transmission) in the idler arm to block out the 869.3-nm emission that arises from the second phase matching condition 
(see Fig.~\ref{anglevswave} and the associated discussion).  We use a long-pass filter (Semrock BLP01-785R-25, $>99.0\%$ transmission) in the signal to block out the 600 
nm arising for the same reason.  We estimate that $\eta_s=34.9\pm1.4\%$ with contributions of $38\pm1.5\%$ from the detector and $8\pm0.2\%$ from the Fresnel reflections from 
the fibers, and $\eta_i=53.3\pm1.4\%$ with contributions of $58\pm1.5\%$ from the detector and $8\pm0.2\%$ from the Fresnel reflections from the fibers.  Using these values, we
 find $\eta_{correct}=68.1\pm4.3\%$, which agrees very well with our predicted value of $71.6\pm1.2\%$  

We measure $R=410\pm12$ Hz, which is corrected using the efficiencies given above to find $R_{correct}=1.89\pm0.19$ kHz, which compares favorably with the predicted joint count 
rate of $1.34\pm0.10$ kHz. 

\textit{Collinear Nondegenerate}  -- Figure~\ref{data}(b) shows the measured and predicted singles spectral rate along the signal beam path, where it is seen that the spectral
 width is much narrower than that observed for degenerate SPDC.  Also, both spectral peaks arise from the two solutions to the phase matching condition as discussed in relation
 to Fig.~\ref{anglevswave}. In the collinear case, the frequencies are conjugates of each other.  The agreement with the predicted spectrum in both relative heights and spectral widths for both spectral features is excellent.

To measure the heralding efficiency, we use use a single lens to collect both the signal and idler beams, couple into single mode fiber, and then back to free space where we use 
a dichroic mirror to separate the signal and idler beams, which are then coupled into multi-mode uncoated fibers and sent to the photon counting detectors.  Because the achromat 
is not perfectly correct, the back-propagated waist at the crystal are different: $W_s=120\pm10$ $\mu$m and $W_i=95\pm10$ $\mu$m.  The pump light is blocked using a high-pass 
filter (Semrock, Di02-R405-25x36 ) that has a transmission at the signal and idler wavelengths exceeding $95\%$ and a pump suppression greater than a factor of 10$^{6}$.  
No narrow-band spectral filters are placed in the signal or idler beam paths.

We find that $\eta=33.5\pm0.5\%$, which is corrected to yield $\eta_{correct}=86\pm5\%$.  Here, we estimate $\eta_s=29.2\pm1.8\%$ with contributions 
contributions of $38\pm1.5\%$ and a combined coupling loss of $77.5\% \pm1\%$ from coupling the beam into freespace and back to fiber and $98\%\pm1\%$ from the filter, 
and $\eta_i=53.3\pm2.3\%$ with contributions  $58\pm1.5\%$ and a combined coupling loss of $92.8\% \pm0.7\%$ and $98\%\pm1\%$ from the filter.
This result is considerably lower than the predicted value of $94.8\pm0.5\%$, 
where we have accounted for the different signal and idler waists in the theory.  One possible reason for the lower measured heralding efficiency
 is that there is a shift in the location of the waist locations of the signal and idler beams of $\sim 250$ $\mu$m.  This non-ideality can be avoided by first spatially 
separating the signal and idler beams and coupling them into independent single-mode fibers.

We measure $R=392\pm12$ Hz for, which is corrected to yield $R_{correct}=2.45\pm0.19$ kHz, while the predicted joint count rate is $1.82\pm0.08$ kHz, where the agreement is 
reasonable.  

\textit{Tighter focusing} -- We also decreased the beam waists for the case collinear nondegenerate configuration, using $W_p = 150\pm5$ $\mu$m, $W_s = 67\pm5$ $\mu$m, and 
$W_{i} = 47\pm5$ $\mu$m.  We find $\eta=32.2\pm0.6\%$, which is corrected to $\eta_{correct}=82.6\pm4\%$, while the theoretical prediction is slightly lowered for these beam parameters
 and is equal to $89\pm2\%$, which is slightly higher than the observations, although still nearly within error.  Finally, we measure $R=890\pm45$ Hz, which is corrected to $R_{correct}=5.53\pm0.62$ kHz, which compares
 favorably with the predicted value of $4.81\pm0.55$ kHz.  

\section{Conclusions and Discussion}
In conclusion, we develop a theoretical formalism to predict the joint spectral rate and the singles spectral rates for Type-I SPDC process using a thin nonlinear optical crystal. 
 From these predictions, the physical factors influencing the heralding efficiency and total joint rate are identified, allowing us to design a system that has an intrinsically 
high efficiency and rate.  We also find that the heralding efficiency depends on the focusing tightness for the noncollinear down-conversion configurations, especially for the 
degenerate case, and that the collinear configurations are less sensitive to this parameter. Finally, we obtain good agreement between theoretical predictions and experimental 
observations.

We compare a few of our results to previous findings that optimize the SPDC heralding efficiency and count rate when the light is coupled into single mode optical fibers.  
Migdall \textit{et al.} (see Fig.~4, Ref. \cite{Migdall}) show the heralding efficiency increases as the signal waist, relative to the pump waist, decreases in the limit of a 
thin crystal, and that the joint count rate decreases with decreasing pump waist for both the collinear and noncollinear regimes. This is in agreement with what our theory predicts
 for pump waist scaling. This tradeoff between heralding efficiency is also in agreement with Bennink \cite{Bennink}, who finds that, as the pump waist is focused tighter, 
the heralding efficiency decreases, and the joint count rate increases. We agree generally with this result, although Bennink considers the thick-crystal regime.  Recently, good 
agreement between experiments and Bennink's predictions has been reported by Dixon \textit{et al.} \cite{Dixon}.  Baek and Kim \cite{Baek} show that both the joint and singles 
spectrum for frequency degenerate, Type-I collinear have a broad bandwidth, while for the noncollinear configuration, the singles spectrum remains broad, but the joint spectrum 
is much narrower, which is consistent with our findings.  Finally, Carrasco \textit{et al.} \cite{Carrasco} show that the singles spectrum can be broadened in both the collinear 
and noncollinear configurations by decreasing the pump waist, and that the joint spectrum for noncollinear can also be broadened by tighter focusing of the pump beam.  Our theory
 agrees with the results presented in Fig. 1 in \cite{Carrasco}.

If we were to replace our detectors with high-quantum efficiency devices, such as recently develop WSi superconducting nanowire detectors with close to 100\% detection 
efficiency \cite{SaeWoo} and replace anti-reflection-coated
 fibers, heralding efficiencies over 80\% are possible for noncollinear degenerate SPDC, and potentially over 90\% for the collinear nondegenerate configuration if we were 
to solve the issue of the chromatic focusing of the signal and idler beams.  Our work paves the way for optimized SPDC sources that will find applications in fundamental quantum 
information science as well as in practical quantum key distribution systems.

\section*{Acknowledgment}

The authors gratefully acknowledge financial support from the DARPA DSO InPho project and for extensive discussion on obtaining high heralding efficiency in SPDC with Paul Kwiat and Bradley Christensen.  

\ifCLASSOPTIONcaptionsoff
  \newpage
\fi

\bibliographystyle{IEEEtran}
\bibliography{bibtry1ieee}

%

\begin{IEEEbiography}[{\includegraphics[width=1in,height=1.25in,clip,keepaspectratio]{guilbert.pdf}}]{Hannah E. Guilbert}
Hannah E. Guilbert (S) was born in Trenton, NJ on Feburary 2, 1987. She recieved the B.A. in Physics from Boston College, Chestnut Hill, MA in 2008.
Ms. Guilbert is currently pursuing the Ph.D. degree in Physics at Duke Univeristy, Durham, NC. Her main focus is in quantum key distribution using nonlinear optics.
\end{IEEEbiography}

\begin{IEEEbiography}[{\includegraphics[width=1in,height=1.25in,clip,keepaspectratio]{GauthierLabIEEEv2.pdf}}]{Daniel J. Gauthier}
Daniel J. Gauthier (M) is the Robert C. Richardson Professor of Physics at Duke University.  He received the B.S., M.S., and Ph.D. degrees from the University of Rochester, 
Rochester, NY, in 1982, 1983, and 1989, respectively. His Ph.D. research on “Instabilities and chaos of laser beams propagating through nonlinear optical media” was supervised 
by Prof. R.W. Boyd and supported in part through a University Research Initiative Fellowship. 

From 1989 to 1991, Prof. Gauthier developed the first CW two-photon optical laser as a 
Post-Doctoral Research Associate under the mentorship of Prof. T. W. Mossberg at the University of Oregon. In 1991, he joined the faculty of Duke University, Durham, NC, 
as an Assistant Professor of Physics and was named a Young Investigator of the U.S. Army Research Office in 1992 and the National Science Foundation in 1993.
He was chair of the Duke Physics Department from 2005 - 2011 and is a founding member of the Duke Fitzpatrick Institute for Photonics. His research interests include:
 high-rate quantum communication, quantum repeaters, nonlinear quantum optics, single-photon all-optical switching, applications of slow light in classical and quantum 
information processing, and synchronization and control of the dynamics of complex networks in complex electronic and optical systems.

 Prof. Gauthier is a Fellow of the
 Optical Society of America and the American Physical Society.
\end{IEEEbiography}


\end{document}